\documentclass[prl,twocolumn,showpacs]{revtex4}
\usepackage{graphicx,amsmath}

\newcommand{\figwidth}{0.7\columnwidth}
\newcommand{\eq}[1]{Eq.(\ref{#1})}
\newcommand{\fig}[1]{Fig.~\ref{#1}}
\newcommand{\avg}[1]{ {\langle #1 \rangle} }
\newcommand{\olcite}[1]{Ref.~\onlinecite{#1}}

\newcommand{\rc}{\rho_{\rm cr}}
\newcommand{\rl}{\rho_{\rm L}}
\newcommand{\rg}{\rho_{\rm G}}
\newcommand{\zc}{z_{\rm c}}
\newcommand{\zp}{z_{\rm p}}
\newcommand{\zpcr}{z_{\rm p,cr}}
\newcommand{\lc}{l_{\rm c}}
\newcommand{\sigmap}{\sigma_{\rm p}}
\newcommand{\nc}{N_{\rm c}}
\newcommand{\np}{N_{\rm p}}
\newcommand{\rhoc}{\rho_{\rm c}}
\newcommand{\pc}{P_L(\rhoc|\zp,\zc)}
\newcommand{\ucr}{U_4^\star}

\newcommand{\ZPCR}{3.877}

\begin{document}

\title{coexistence diameter in two-dimensional colloid-polymer 
mixtures}

\author{R. L. C. Vink and H. H. Wensink}

\affiliation{Institut f\"ur Theoretische Physik II, Heinrich Heine
Universit\"at D\"usseldorf, Universit\"atsstra{\ss}e 1, 40225
D\"usseldorf, Germany}

\date{\today}

\begin{abstract}

We demonstrate that the {\it law of the rectilinear coexistence diameter} 
in two-dimensional (2D) mixtures of {\it non-spherical} colloids and 
non-adsorbing polymers is violated. Upon approach of the critical point, 
the diameter shows logarithmic singular behavior governed by a term $t \ln 
t$, with $t$ the relative distance from the critical point. No sign of a 
term $t^{2\beta}$ could be detected, with $\beta$ the critical exponent of 
the order parameter, indicating a very weak or absent Yang-Yang anomaly. 
Our analysis thus reveals that non-spherical particle shape alone is {\it 
not} sufficient for the formation of a pronounced Yang-Yang anomaly in the 
critical behavior of fluids.

\end{abstract}


\pacs{05.70.Jk, 64.60.Fr, 64.70.Fx}

\maketitle

Critical phenomena have been studied extensively for decades. 
Nevertheless, a few important issues remain controversial. One 
longstanding problem concerns the critical behavior of the so-called 
coexistence diameter of fluids undergoing a phase transition from liquid 
to gas (the coexistence diameter is defined as the average density of the 
two coexisting phases). In 1964, Yang and Yang showed that the divergence 
of the constant-volume specific heat at the critical point implies that 
{\it either} $d^2 p /d T^2$ or $d^2 \mu / d T^2$ or {\it both} diverge 
\cite{yang.yang:1964}. Here, $T$ is the temperature, $p$ the pressure, and 
$\mu$ the chemical potential. A remarkable consequence, realized only 
recently by Fisher and co-workers \cite{fisher.orkoulas:2000}, is that the 
divergence of $d^2 \mu / d T^2$ implies that the coexistence diameter 
$\delta$ gains an additional term $t^{2\beta}$ which, assuming Ising 
universality, dominates the previously recognized term $t^{1-\alpha}$ 
\cite{widom.rowlinson:1970, mermin:1971}. As usual, $t$ is the relative 
distance from the critical point; $\alpha$ and $\beta$ are the critical 
exponents of the specific heat and the order parameter, respectively. The 
coexistence diameter near the critical point thus reads as
\begin{equation}\label{eq:coex}
  \delta_{\alpha \neq 0} = \rc \left( 1 + A_{2\beta} t^{2\beta}
	+ A_{1-\alpha} t^{1-\alpha} + A_1 t \right),
\end{equation}
with $\rc$ the critical concentration, and non-universal amplitudes $A_i$. 
The divergence of $d^2 \mu / d T^2$ is called a Yang-Yang (YY) anomaly, 
and $A_{2\beta} \neq 0$ if one is present. For {\it symmetric} fluids, 
such as the Widom-Rowlinson mixture \cite{widom.rowlinson:1970}, $d^2 \mu 
/ d T^2$ remains finite, in which case $A_{2\beta}=0$ \cite{vink:2006}. 
However, realistic fluids are typically {\it asymmetric}, in which case a 
YY-anomaly cannot be ruled out.

Indeed, evidence for a YY-anomaly was found in simulations of asymmetric 
three-dimensional (3D) fluids \cite{orkoulas.fisher.ea:2001, 
kim.fisher.ea:2003}. Experiments on propane and carbon dioxide also show 
evidence for a weak YY-anomaly \cite{orkoulas.fisher.ea:2000}, while for 
He-3 the situation is less clear \cite{anisimov.zhong.ea:2004, 
hahn.weilert.ea:2004}. However, for 3D fluids, the analysis is extremely 
difficult. Assuming 3D Ising universality with $\alpha \approx 0.109$ and 
$\beta \approx 0.326$ \cite{fisher.zinn:1998}, there are two singular 
terms in \eq{eq:coex} which need to be distinguished not only from each 
other, but also from the leading analytic background term $A_1 t$. A 
direct observation of the term $t^{2\beta}$ is consequently very 
difficult. Only very recently, after it was recognized that the amplitudes 
$A_{1-\alpha}$ and $A_1$ are coupled, could both singular terms in 
\eq{eq:coex} be resolved from experimental data \cite{anisimov.wang:2006}. 

Nevertheless, due to the competition between terms in \eq{eq:coex}, 
investigations of the coexistence diameter in 3D fluids remain 
challenging. An attractive alternative, where the competition between 
singular terms is less severe, is to consider a fluid where the specific 
heat diverges logarithmically at the critical point, implying $\alpha=0$. 
The critical behavior of the diameter is then given by
\begin{equation}\label{eq:coex2d}
  \delta_{\alpha=0} = \rc \left( 1 + A_{2\beta} t^{2\beta}
	+ A_0 \, t \ln t + A_1 t \right).
\end{equation}
In other words, the term $t^{1-\alpha}$ is replaced by a (much weaker) 
{\it logarithmic} singularity \cite{anisimov:2006}. In order to detect the 
YY-anomaly, one thus needs to distinguish power law behavior from 
logarithmic behavior, which is expected to yield a more pronounced 
numerical signature. To this end, one could consider a 2D fluid, where 
$\alpha=0$ and $\beta=1/8$ (assuming 2D Ising universality). Note that 
mean-field systems do not qualify, despite having $\alpha=0$. For 
mean-field systems, $\alpha=0$ corresponds to a finite discontinuity in 
the specific heat, yielding a purely rectilinear diameter 
\cite{anisimov:2006}. Still, even though the numerical analysis of a 2D 
fluid with Ising critical behavior may become more simple, this is no 
guarantee that a YY-anomaly will be found. The absence or presence of a 
YY-anomaly is a non-universal feature: $A_{2\beta}$ may well be zero! It 
is not completely clear which features in a fluid determine the strength 
of the YY-anomaly. Obviously, the fluid must be asymmetric. In addition, 
there are experimental indications that molecular shape and symmetry, in 
particular departures from spherical form, are important contributing 
factors \cite{orkoulas.fisher.ea:2000}. From these considerations, it 
appears that the ``minimal'' fluid in which a YY-anomaly may most easily 
be found should (1) be two-dimensional and exhibit a 2D Ising critical 
point (2) be asymmetric, and (3) contain non-spherical particles.

In this work, we will investigate a fluid with precisely these properties, 
and focus on the critical behavior of its diameter. We consider a 2D 
version of the colloid-polymer model of Asakura and Oosawa (AO) 
\cite{asakura.oosawa:1954}, but generalized to non-spherical colloids. In 
the original AO model, colloids and polymers are treated as spheres in 3D, 
assuming hard-core interactions between colloid-colloid and 
colloid-polymer pairs, while polymer-polymer pairs can interpenetrate 
freely (the AO model is thus clearly asymmetric). Since the polymers may 
overlap freely, their translational entropy is increased significantly 
when the colloids group together. Hence, there is an effective (depletion) 
attraction between the colloids. Provided polymer concentration and size 
are sufficiently large, the attraction is strong enough to drive phase 
separation in the AO model, whereby the system splits-up into a 
colloid-rich (polymer-poor) phase, and a colloid-poor (polymer-rich) 
phase. As expected for systems with short-ranged interactions, the 
corresponding unmixing critical point belongs to the 3D Ising universality 
class \cite{vink.horbach:2004*1}. Hence, it is anticipated, although one 
should plan to check, that the AO model in 2D will exhibit a 2D Ising 
critical point. The effect of non-spherical particle shape is incorporated 
by modeling the colloids not as spheres, but as line segments. Our minimal 
model is thus a 2D mixture of colloidal line segments of length $\lc$, and 
effective polymer disks of diameter $\sigmap$, interacting via AO 
potentials. In other words, overlaps between line segments, as well as 
overlaps between line segments and polymer disks are forbidden, while the 
polymer disks may overlap freely.

The aim of this work is to check if a YY-anomaly in this model can be 
found. We will do so using computer simulation and finite size scaling 
(FSS) in the grand canonical ensemble. In this ensemble, the total area of 
the system $A$, the temperature $T$, and the colloid (polymer) fugacity 
$\zc$ ($\zp$) are fixed, while the number of colloids ($\nc$) and polymers 
($\np$) fluctuates. The thermal wavelength is set to unity, such that 
$\zp$ reflects the average concentration $\np/A$ a pure phase of polymers 
would have (recall that such a phase is simply an ideal gas). The 
remaining lengths are expressed in units of $\lc$. The colloid-to-polymer 
size ratio is set to $\sigmap / \lc = 0.95$. At the coexistence colloid 
fugacity, it is expected that the mixture phase separates into a colloid 
poor phase (the gas) and a colloid rich phase (the liquid), provided the 
polymer fugacity exceeds a critical value $\zpcr$. The phase separation is 
thus driven by $\zp$, which therefore plays a role analogous to inverse 
temperature in gas-liquid transitions of simple fluids. The relative 
distance from the critical point is written as $t \equiv \zp/\zpcr-1$, and 
$\rg$ ($\rl$) denotes the concentration $\rhoc \equiv \nc/A$ of colloids 
in the gas (liquid) phase. A natural order parameter is the density gap 
$\Delta \equiv (\rl-\rg)/2$, while the coexistence diameter reads as 
$\delta \equiv (\rl+\rg)/2$. In the limit $t \to 0$, we expect critical 
power law behavior $\Delta \propto t^\beta$ for the order parameter, and 
\eq{eq:coex2d} for the coexistence diameter.

\begin{figure}
\begin{center}
\includegraphics[clip=,width=\figwidth]{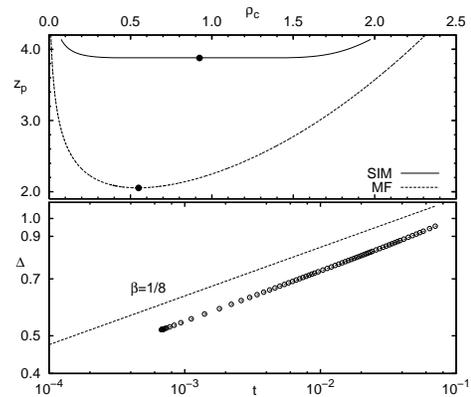}

\caption{\label{phase} {\it upper frame:} Binodals of the 2D 
colloid-polymer mixture of this work, obtained using mean-field theory 
(MF) and simulation (SIM); circles mark critical points. {\it lower 
frame:} Order parameter $\Delta$ as function of $t$, obtained using 
simulation and FSS; a value $\zpcr = \ZPCR$ in $t$ was used.}

\end{center}
\end{figure}

For our minimal model, the critical point must be located first, in 
particular the value of $\zpcr$. Next, 2D Ising universality of the 
critical point must be established, after which the diameter can be 
investigated. To ``guide'' the simulations, the phase diagram is obtained 
approximately first, using a simple mean-field (MF) theory based on a 
free-volume approach for 2D rod-polymer mixtures. The free-volume fraction 
and the free energy for a pure system of 2D lines are derived from scaled 
particle theory \cite{lekkerkerker.stroobants:1994}. The free energy is 
exactly the same as the one obtained within Onsager's second virial theory 
\cite{lekkerkerker.stroobants:1994}. The resulting binodal is shown in 
\fig{phase}. The theory predicts the critical point at $\rc^{\rm MF} 
\approx 0.550$ and $\zpcr^{\rm MF} \approx 2.054$. No nematic ordering of 
the colloids is predicted in the direct vicinity of the critical point: 
the coexisting phases are isotropic. Since the theory ignores critical 
fluctuations, it is expected that $\zpcr^{\rm MF}$ underestimates the true 
value $\zpcr$ significantly. Nevertheless, the theoretical result is 
important because it provides an indication in which regime the (time 
consuming) simulations need to be carried out. The simulations are 
performed in the grand canonical ensemble, on a 2D square of size $A = L 
\times L$ using periodic boundary conditions. We measure the distribution 
$\pc$, defined as the probability of observing a system with colloid 
concentration $\rhoc$, at fugacities $\zp$ and $\zc$, with $L$ the system 
size. The insertion and removal of particles is performed using a cluster 
move \cite{vink.horbach:2004*1}, combined with a biased sampling scheme 
\cite{virnau.muller:2004} to overcome the free energy barrier separating 
the phases, and histogram reweighting \cite{ferrenberg.swendsen:1989}. To 
obtain a single distribution, around 100 CPU hours for a small system 
($L=22$), and 350~h for a large system ($L=30$), are required.

\begin{figure}
\begin{center}
\includegraphics[clip=,width=\figwidth]{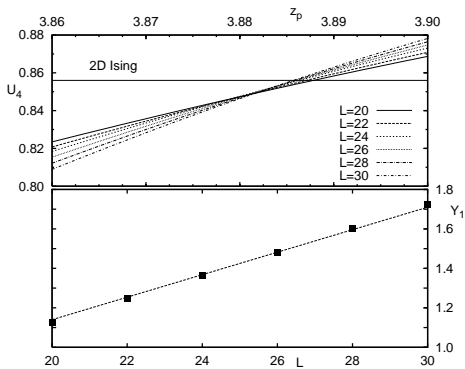}

\caption{\label{cumulant} Cumulant analysis near the critical point. The 
top frame shows $U_4$ as function of $\zp$ for various system sizes $L$ as 
indicated. The lower frame shows the cumulant slope $Y_1$ at $\zpcr$ as 
function of $L$.}

\end{center}
\end{figure}

\begin{figure}
\begin{center}
\includegraphics[clip=,width=\figwidth]{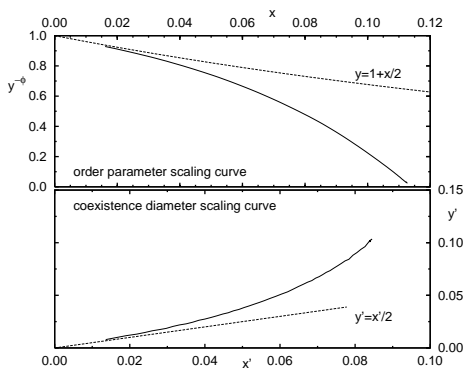}

\caption{\label{scaling} {\it upper frame:} Order parameter scaling curve 
$y=f(x)$ (solid line). Following convention \cite{kim.fisher.ea:2003}, the 
scaling curve is raised to a negative exponent with $\phi = 1/\beta$, 
where the 2D Ising value $\beta=1/8$ was used. Also shown is the small $x$ 
limiting form $y=1+x/2$ (dashed line). {\it lower frame:} Coexistence 
diameter scaling curve $y'=g(x')$ (solid line). The dashed line shows the 
small $x'$ limiting form $y'=x'/2$.}

\end{center}
\end{figure}

A standard route to obtain $\zpcr$ in simulations, is to measure the 
$L$-dependence of the cumulant $U_4 = \avg{m^2}^2 / \avg{m^4}$ along some 
path in the $(\rhoc,\zp)$-plane. Here, $m = \rhoc - \avg{\rhoc}$, and 
$\avg{\cdot}$ denote grand canonical averages. The cumulant becomes 
system-size independent at the critical point \cite{binder:1981}. Plots of 
$U_4$ as function of $\zp$ for different system sizes $L$ are expected to 
show a common intersection point, leading to an estimate of $\zpcr$. 
Moreover, the value of the cumulant $\ucr$ at the intersection point is 
universal, and this gives an indication of the universality class. The 
result is shown in the upper frame of \fig{cumulant}. The data were 
obtained using the colloid fugacity at which $\avg{m^2}$ is maximized 
\cite{orkoulas.fisher.ea:2001}. From the intersections, we obtain $\zpcr = 
3.881 \pm 0.005$, where the error reflects the scatter in the various 
intersection points. At the intersection point $\ucr \approx 0.85$, which 
is very close to the accepted 2D Ising value $\ucr \approx 0.856$ 
\cite{kamieniarz.blote:1993} (horizontal line in \fig{cumulant}). While 
this already suggests 2D Ising universality, additional confirmation is 
obtained from the critical exponents. We consider $\beta$ and $\nu$, with 
$\nu$ the critical exponent of the correlation length. Here, $\nu$ is 
extracted from the cumulant slope $Y_1 \equiv d U_4 / d \zp$ {\it at} the 
critical value of $\zp$. One expects that $Y_1 \propto L^{1/\nu}$, with 
$L$ the system size. The lower frame of \fig{cumulant} shows $Y_1$ as 
function of $L$, where the above estimate of $\zpcr$ was used. The line is 
a linear fit through the origin, which describes the data very well, and 
thus confirms the 2D Ising value $\nu=1$. To obtain $\beta$, we apply the 
FSS algorithm of \olcite{kim.fisher.ea:2003}, using system sizes 
$L=20-30$. Starting with $\zp$ significantly {\it above} its 
critical value, the algorithm proceeds by plotting $U_4$ as function of 
the average colloid concentration $\avg{\rhoc}$. The resulting plot 
reveals two minima, located at $\rho^-$ and $\rho^+$, with respective 
values $Q^-$ and $Q^+$ at the minima. Defining the quantities $Q_{\rm min} 
= (Q^+ + Q^-)/2$, $x = Q_{\rm min} \ln (4/e Q_{\rm min} )$, and $y = 
(\rho^+ - \rho^-) / (2 \Delta)$, the points $(x,y)$ obtained for different 
system sizes $L$ should, in the limit far away from the critical point, 
collapse onto the line $y=1+x/2$. Recall that $\Delta$ is the order 
parameter in the thermodynamic limit at the considered fugacity $\zp$, 
precisely the quantity of interest, which may thus be obtained by fitting 
until the best collapse onto $1+x/2$ occurs. In the next step, $\zp$ is 
chosen closer to the critical point, the points $(x,y)$ are calculated as 
before, but this time $\Delta$ is chosen such that the new data set 
joins smoothly with the previous one, yielding an estimate of the order 
parameter at the new fugacity. This procedure is repeated as closely as 
possible to the critical point, where $\Delta$ vanishes. The output of the 
algorithm, $\Delta$ as function of $\zp$, is then fitted to $\Delta 
\propto t^\beta$, in order to estimate $\zpcr$ and $\beta$. We obtain 
$\zpcr = \ZPCR \pm 0.001$, which is consistent with the (less precise) 
analysis of \fig{cumulant}. Shown in the lower frame of \fig{phase} is the 
order parameter as function of $t$, on double logarithmic scales. The line 
has slope $\beta=1/8$, and confirms the 2D Ising exponent in the 
simulation data. The FSS algorithm also yields the order parameter scaling 
curve $y=f(x)$ \cite{kim.fisher.ea:2003}, with $x$ and 
$y$ defined as above, shown for completeness in the upper frame of 
\fig{scaling}. Away from the critical point ($x \to 0$), the scaling curve 
has the limiting form $y=1+x/2$; at the critical point, the scaling curve 
diverges. The significance of the scaling curve is its universal 
character: all systems with a 2D Ising critical point should yield a 
scaling curve for the order parameter similar to the one shown here 
\cite{kim.fisher.ea:2003}.

\begin{figure}
\begin{center}
\includegraphics[clip=,width=\figwidth]{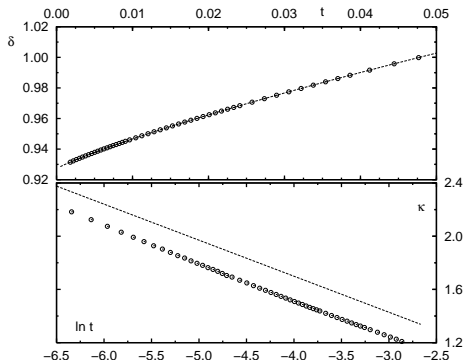}

\caption{\label{diam} {\it upper frame:} Coexistence diameter $\delta$ as 
function of $t$ (circles). The dashed line is a fit to \eq{eq:coex2d}. 
{\it lower frame:} $\kappa$ as function of $\ln t$ (circles). The straight 
line (dashed) confirms the logarithmic nature of the divergence. In both 
plots, $\zpcr=\ZPCR$ in $t$ was used.}

\end{center}
\end{figure}

At this point, sufficient evidence for 2D Ising universality has been 
provided. Since the order parameter is a scalar, and since the 
interactions are short-ranged, there are in any case no theoretical 
grounds to contemplate a different universality class. Hence, we will now 
consider the coexistence diameter. To obtain the diameter, the FSS 
algorithm of \olcite{kim.fisher.ea:2003} is used. Similarly, also for the 
coexistence diameter, a scaling curve $y'=g(x')$ is generated, based on 
{\it different} quantities $x'$ and $y'$ defined in 
\olcite{kim.fisher.ea:2003}. In contrast to the order parameter, the 
scaling curve of the diameter is {\it not} universal. For our model, the 
corresponding curve is shown in the lower frame of \fig{scaling}. For 
small $x'$, it correctly approaches the exact limiting form $y'=x'/2$ 
\cite{kim.fisher.ea:2003}. The curvature at $x' \gg 0$ already suggests 
singular behavior. According to \eq{eq:coex2d}, this may reflect the 
YY-anomaly, or logarithmic behavior, or both. To quantify this, the 
diameter itself is shown in the upper frame of \fig{diam}. The quality of 
the data is such that $\delta$ can be resolved down to $t \approx 0.0015$. 
A fit to \eq{eq:coex2d} yields $\rc = 0.9270 \pm 0.0006$, $A_{2\beta} 
\approx 0$, $A_0=-0.29 \pm 0.01$ and $A_1 = 0.76 \pm 0.03$, where the 
error reflects the scatter resulting from the range over which the fit is 
performed. The fit describes the data perfectly well, without the need for 
a term $t^{2\beta}$, indicating {\it logarithmic} singular behavior. In 
other words, despite the non-spherical particle shape in our model, the 
present analysis does {\it not} reveal a YY-anomaly. For completeness, in 
the lower frame of \fig{diam}, the derivative $\kappa \equiv d\delta/dt$ 
is plotted as function of $\ln t$. In case of singular behavior, $\kappa$ 
is expected to diverge as $t \to 0$. The (logarithmic) divergence is 
clearly visible. Finally, by combining the coexistence diameter and order 
parameter data, the binodal in the thermodynamic limit was constructed, 
see the top frame of \fig{phase}.

In conclusion, we have shown that 2D colloid-polymer mixtures, with 
non-spherical colloids, do not display a pronounced YY-anomaly, at least 
not for the colloid-to-polymer size ratio $q=0.95$ considered by us. 
Although the diameter becomes singular upon approach of the critical 
point, the singularity is logarithmic, and well described by the 
theoretically expected term $t \ln t$. While it has been suggested 
\cite{orkoulas.fisher.ea:2000} that non-spherical particle shape may be an 
important contributing factor to the formation of the YY-anomaly, this 
seems not to be the case for our 2D model. If a YY-anomaly is present in 
our 2D model nevertheless, it is very weak, and negligible down to $t 
\approx 0.0015$ accessible in our simulations. In contrast, our results 
may be compatible with the very recent \olcite{anisimov.wang:2006}, where 
it is argued that a YY-anomaly is expected when $\rho^\star = \rc a_{\rm 
i}$ is small, where $a_{\rm i}$ represents the typical interaction volume. 
For our model, $a_{\rm i} \approx q^2$, and so $\rho^\star \approx 0.8$. 
This value even exceeds $\rho^\star \approx 0.75$ of the Widom-Rowlinson 
mixture \cite{widom.rowlinson:1970}, for which no YY-anomaly was detected 
either \cite{vink:2006}. In order to detect a YY-anomaly in our 2D model, 
it seems that smaller size ratios $q$ are required; this could be a topic 
for further simulations. It is tempting to speculate if the 2D model 
considered in this work can also be realized experimentally. Colloidal 
particles, due to their mesoscopic size, pose many advantages over atomic 
fluids. This has already enabled the investigation of critical phenomena 
in 3D, whereby the particles are visualized directly using confocal 
microscopy. Other applications may be found in order-disorder phase 
transitions in adsorbed monolayers of atoms or small molecules at 
surfaces.

\acknowledgments

This work was supported by the {\it Deutsche Forschungsgemeinschaft} under 
the SFB-TR6/D3. HHW also acknowledges the {\it Alexander von Humboldt} 
foundation. We thank M. A. Anisimov and K. Binder for valuable comments.

\bibstyle{revtex}
\bibliography{mainz}

\end{document}